\documentclass[prd,twocolumn,epsfig]{revtex4}

\newcommand{\be}{\begin{equation}}
\newcommand{\ee}{\end{equation}}
\newcommand{\<}{\langle}
\renewcommand{\>}{\rangle}

\newcommand{ \K }{ \raisebox{-0.8ex} {\scriptsize \it K} }

\begin{document}

\title{Boson Dominance in nuclei}

%\author{Fabrizio Palumbo~\thanks{This work has been partially 
%  supported by EEC under the contract HPRN-CT-2000-00131}}
%\address{INFN -- Laboratori Nazionali di Frascati - P.~O.~Box 13, I-00044 Frascati, ITALIA}

\author{Fabrizio Palumbo}
 \altaffiliation[]{This work has been partially 
  supported by EEC under the contract HPRN-CT-2000-00131}%Lines break automatically or can be forced with} \\
\affiliation{%
  {\small\it INFN -- Laboratori Nazionali di Frascati}  \\%[-0.2cm]
  {\small\it P.~O.~Box 13, I-00044 Frascati, ITALIA}          \\%[-0.2cm]
  {\small e-mail: {\tt fabrizio.palumbo@lnf.infn.it}}     
}%

%\author{
%  { Fabrizio Palumbo~\thanks{This work has been partially 
%  supported by EEC under the contract HPRN-CT-2000-00131}}             \\%[-0.2cm]
%  {\small\it INFN -- Laboratori Nazionali di Frascati}  \\%[-0.2cm]
%  {\small\it P.~O.~Box 13, I-00044 Frascati, ITALIA}          \\%[-0.2cm]
%  {\small e-mail: {\tt fabrizio.palumbo@lnf.infn.it}}     
%   }

%\date{\today}

\thispagestyle{empty}   % Suppress page number on front page.

\begin{abstract}

 We present a  new method of bosonization of fermion systems applicable when the partition function is
dominated by composite bosons. Restricting the partition function to such states we get an euclidean bosonic
action from which we derive the Hamiltonian. Such a procedure respects all the fermion symmetries, in particular
fermion number conservation, and provides a boson mapping of all fermion operators.

%\vspace{0.3cm}

%\noindent Keywords:

\end{abstract}

\maketitle

\clearpage

%\nopagebreak

\section{Introduction}

The importance of effective bosons in nuclear physics became clear after the observation that heavy deformed nuclei
display some of the features of superconducting systems~\cite{Bohr}. In these nuclei there must then be Cooper pairs
of nucleons. In this spirit Arima and Iachello introduced~\cite{Arim} two different composite bosons, the
$s$- and $d$-bosons. Their model, the  Interacting Boson Model, proved extremely successful in reproducing low energy
nuclear properties, but has not yet  been derived in a fully satisfactorily way from a nuclear Hamiltonian.

Many attempts to reformulate the nuclear Hamiltonian in terms of effective bosons, however, were done before
the Interacting Boson Model was invented. Of special importance are the works of Beliaev and
Zelevinsky~\cite{Beli}, who constructed a composite boson operator requiring its commutation with the nuclear
Hamiltonian, and of Marumori, Yamamura and Togunaga~\cite{Maru}, who developed a method based on a 
map of fermion into boson matrix elements.  

The first important step in the derivation of the IBM respecting nucleon number conservation is due to
Otsuka, Arima and Iachello~\cite{Otsu}. Their work is based on a map of a single $j$-shell nucleon space into a
 boson  space. The boson Hamiltonian so obtained reproduces  exactly the spectrum of the pairing model. Their procedure 
has been somewhat extended~\cite{Otsu1} but not sufficiently generalized. 

There are several recipes for bosonization~\cite{Klei} based on a mapping of the nucleon model
space into a boson space. Such methods do not violate nucleon conservation and in principle yield an exact
solution to the problem, but in practice one has to perform a truncation in the nucleon space related to a selection of
 degrees of freedom guided by physical insight and calculational convenience. One shortcoming of this procedure
is the appearance of "intruders", namely states which in spite of their low energy do not appear in the boson space
generated by the mapping~\cite{Diep}.

A different approach to bosonization which does not require a preliminary truncation of the nucleon space
and does not violate nucleon conservation~\cite{Namb,Barb} is based on the Hubbard-Stratonovich transformation. The
latter renders quadratic the fermion interaction by introducing bosonic auxiliary fields which in the end become the
physical fields. The typical resulting structure is that of chiral theories~\cite{Mira}. In such an approach an energy
scale emerges naturally, and only excitations  of lower energy can be described by the auxiliary fields. 
In our opinion  this approach has not received enough attention and its potentiality has not been fully explored.

The physical idea behind bosonization is that certain composite bosons dominate the partition function at low energy
(Boson Dominance), an assumption certainly justified for Goldstone bosons like Cooper pairs. We present a new way to
implement Boson Dominance. We introduce generic nucleon composites whose structure  will be determined at the end
by a variational procedure, and evaluate the partition function restricted to such composites. In this way we  get an
euclidean bosonic action in closed form.

In the derivation of the effective action we need only one approximation, concerning the identity in the space of the
composites, but we respect all the nucleon symmetries, in particular nucleon number conservation.  We 
emphasize  that the  closed form of the action opens the way to numerical simulations of fermion systems in terms of
bosonic variables, avoiding the "sign problem"~\cite {Kalo}.

Bosonization is achieved within the path integral formalism. In this framework the standard procedure to evaluate
physical quantities is to first find the minimum of the action at constant fields. Depending on the solution, one has
spherical or deformed nuclei. In the latter case rotational excitations appear as Goldstone modes associated to the
spontaneous breaking of rotational symmetry. The notion of spontaneous symmetry breaking survives in fact with a
precise definition also in finite systems~\cite{Barb}. Next the quantum fluctuations must be taken into account.
But we anticipate a subtlety in this program which in the present case is not completely standard due to the composite
nature of the bosons.

In nuclear physics the Hamiltonian formalism is of much wider application. Since the effective bosonic action,
due to compositness, is not in canonical form, it has been necessary to devise an appropriate  procedure to derive
the Hamiltonian. In this context the mentioned subtlety finds a natural solution. 

Only composites which have components on many nucleon states can be approximated by bosons. But the boson space
cannot be arbitrarily truncated. For instance, even if the nuclear potential contains only monopole and quadrupole
pairing interactions, the $s$- and $d$-bosons will be coupled to all the other bosons permitted by angular momentum
conservation. An approximate decoupling should arise dynamically, but at the moment a clearcut mechanism 
is not known. We will say more about this point in our Conclusions.    
 
Bosonization appears in several many-fermion systems and relativistic field theories. The effective bosons fall into
two categories, depending on their fermion number. The Cooper pairs of the BCS model of superconductivity, of the
Interacting Boson Model of Nuclear Physics, of the  Hubbard model of high $T_c$ superconductivity~\cite{Cini} and of
color superconductivity in QCD have fermion number 2. Similar  composite bosons with fermion number zero appear as
phonons, spin waves and chiral mesons in QCD. The latter bosons can be included in the present formalism by replacing
in the composites one fermion operator by an antifermion (hole) one. This becomes necessary when the interaction
contains, as in nuclear physics, important particle-hole terms. This extension of the method will be presented
in a future work.

A preliminary presentation of our results was given in \cite{Palu}.

The paper is organized as follows.
In Section II we outline  our approach. In Section III we report the bosonic effective action. In Section IV we  derive
the bosonic Hamiltonian as  the normal ordered form of a nonpolynomial function of creation-annihilation boson
operators. This result includes the boson mapping of all fermion operators. For a practical use this Hamiltonian must be
expanded in the inverse of the dimension of the nucleon space.  In Section V we report an independent derivation
of the boson Hamiltonian valid for a small number of nucleons and we specify this Hamiltonian
to the case  a single $j$-shell. In Section VI come back to the path integral
formalism, introducing the Goldstone and Higgs fields, and in Section VII we end with our conclusions.

In the presentation of our results, to facilitate the understanding of the logical development, we choose to
relegate  many technical details in a number of Appendices. In Appendix \ref{A} we reported some results
concerning Berezin integrals and in Appendix \ref{B} their use in calculations with coherent states of composites.
The basics of this formalism can be found in a condensed form in ~\cite{Nege}, while an exhaustive presentation is
given in~\cite{Ripk}. In Appendix \ref{C} we discussed the properties of an operator which approximates the identity in
the space of the composites and in Appendix \ref{D} some intermediate steps in the derivation of the effective bosonic
action.

\section{Outline of the approach}

Consider a nuclear partition function 
\be
Z = \mbox{tr} \left[ \exp \left( - { 1 \over T} ( H - \mu_N \hat{ n}_N) \right) \right]
\ee
where
$T$ is the temperature,  $\mu_N$ the nucleon chemical potential and $\hat{ n}_N $ the nucleon number operator.
 A sector of $n_N$ nucleons can be selected by the constraint
\be
T {\partial \over \partial \mu_N} \ln Z = n_N.             \label{number}
\ee
Under the assumption of Boson Dominance we can restrict the trace to nucleon bosonic composites. The restricted
partition function can be written
 \be
Z_C = \mbox{tr} \left[ {\mathcal P} \exp \left( - { 1 \over T} (H- \mu_N \hat{ n}_N)  \right) \right]
\ee
where ${\mathcal P}$ is a projection operator in the subspace of the composites. We will only be able to construct 
an approximation to such operator. This is the only approximation we will do (beyond the physical assumption of Boson
Dominance). In analogy to the case of elementary bosons we will assume
\be 
{\mathcal P} = \int d \mu (\beta^*,  \beta )   \, |\beta \>\<\beta|  \label{Poperator}
\ee
where $  d \mu (\beta^*,  \beta ) $ is an integration measure (to be specified later) over the holomorfic variables
$\beta^*, \beta$ and the $|\beta \>$ are coherent states of composites
\be
|\beta \> = |\exp \left( \sum_J \beta^*_J {\hat b}^{\dagger}_J \right) \>.
\ee
Definition and properties of the operator $\mathcal{ P}$ are discussed in Appendix \ref{C}. The $|\beta \> $
 are defined in terms of composite creation operators  
\be
{\hat b}^{\dagger}_J = { 1 \over 2 \sqrt{\Omega_J}} c^{\dagger} B^{\dagger}_J c^{\dagger}
= { 1\over 2\sqrt{\Omega_J}} \sum_{m_1,m_2}c^{\dagger}_{m_1}\left( B^{\dagger}_J\right)_{m_1,m_2}
 c^{\dagger}_{m_2}.
\ee
The $c^{\dagger}$'s are nucleon creation operators, the $m$'s represent all the nucleon  quantum numbers, the
matrices $B_J$ are the form factors of the composites with  quantum numbers $J$.  $\Omega_J$ is the index of
nilpotency of the $J$-composite, which is defined as the largest integer such that
\be
\left( \hat{b}_J \right)^{\Omega_J} \neq 0. 
\ee
In the present paper we will assume for simplicity the index of nilpotency independent of the quantum numbers
of the composites, and equal to half the dimension of the nucleon space, but we will mention a possible consequence
of this simplification.

It is obvious that a necessary condition for a composite to resemble an elementary boson, is that its index of
nilpotency  be large. But such condition is in general not sufficient. Consider for instance the case
\be
\left( B^{\dagger} B \right)_{m_1 m_2} = \delta_{m_1 m_2} d_{m_1},
\ee
where $d_1=1$, $d_m \ll 1, m \neq 1$. Such a composite, irrespective of its index of nilpotency, consists essentially
of a unique  state of a nucleon pair. We must instead require that the composites actually live in a large part of
the  nucleon space. This can be ensured by the further requirement
\be
\det \left( B_J^{\dagger} B_J \right) \sim  1.  \label{cond}
\ee
Solutions to the equations for the $B$-matrices which do not satisfy the above condition must be discarded.

Evaluation of the trace (reported in Appendix \ref{D}) gives
\be
Z_C=\int \left[ { d \beta^* d \beta \over 2 \pi i} \right]\exp \left( - S_{\mbox{eff}}(\beta^*, \beta) \right).
\ee

 Bosonization is thus achieved, and the nuclear dynamics can be studied by functional or numerical methods. The last
possibility appears interesting because it avoids the "sign" problem \cite{Kalo} which affects the Monte Carlo approach
to the study of many-fermion systems..

In nuclear physics it is generally used  a Hamiltonian formalism. The Hamiltonian of the effective bosons, $H_B$,
cannot be read directly from the effective action, because $S_{\mbox{eff}} (\beta^*,\beta)  $ does not have the form
of an action of elementary bosons. Indeed it contains anomalous time derivative terms, anomalous couplings of the
chemical potential and nonpolynomial interactions, which are all features of compositeness. Therefore it has
been necessary to devise an appropriate procedure to derive $H_B$, which also has been given in closed form,
in terms of boson operators $b^{\dagger}, b$, satisfying canonical commutation relations. In conclusion
\be
Z_C = \mbox{tr} \left( -{ 1 \over T} (H_B - \mu_B \hat{ n}_B) \right)
\ee
where $\mu_B$ is the boson chemical potential and $\hat{ n}_B $ the boson number operator.

For a practical use, however, it is necessary to perform an expansion of $H_B$. The expansion parameter is the
inverse of $\Omega$.

\section{The effective bosonic action}

 We write the nucleon-nucleon potential as a sum of multipole pairing terms, so that the Hamiltonian has the form
\begin{eqnarray}
H &=& c^{\dagger} h_0 \, c -
\sum_K g\K  \, { 1\over 2} \, c^{\dagger} F_K^{\dagger} c^{\dagger} \, 
{ 1\over 2}\, c \, F_K \, c. \label{Hami}
\end{eqnarray}
 The one body term includes the single-particle energy 
with matrix $e$, the nucleon chemical potential $\mu_N$ and any single particle interaction with external fields
included in the matrix ${\cal M}$  
\be
h_0= e - \mu_N + {\cal M} .
\ee
The matrices $F_K $ are the form factors of the potential, normalized according to
\be
 \mbox{tr} ( F_{K_1}^{\dagger} F_{K_2}) = 2 \, \Omega \delta_{K_1 K_2}.
\ee
Any potential can be written in the above form~\cite{Loiu}. But this form is not convenient, as it is well known, when 
particle-hole terms are important. To properly account for such terms in the present scheme it is necessary to introduce
phonons, which will be done in a separate work.

In order to evaluate $Z_C$ we divide the inverse temperature in $N_0$ intervals of size $\tau$
\be 
T = { 1 \over N_0 \tau}.
\ee
Then as shown in Appendix D the  euclidean effective action has the form
 \begin{eqnarray}
& & S_{\mbox{eff}}(\beta^*,\beta) = N_0 \ln \mathcal{ J} +\tau \sum_t \mbox{tr} \left\{ { 1\over 2 \tau}   \ln \left[
1\!\!1 + 
 \tau \, \Gamma \, \Phi^* \nabla_t \Phi \right] \right.
\nonumber\\
& & \left. - { 1\over 4}  \sum_K g_K  \left[   (\Gamma \Phi^*  F_K^{\dagger}) \,
\mbox{tr}(\Gamma F_K \Phi) 
 +  2 \, \left( (\Gamma -1) \,  F^{\dagger}_K F_K \right)
\right. \right.
\nonumber\\
& & \left. \left.
-  [ \Gamma \Phi^* F_K^{\dagger}, \Gamma  F_K \Phi]_+  \right]
 + { 1\over 2}   \left[ \Gamma  \, \Phi^*  
\, ( \Phi \, h^T + h \,\, \Phi ) \right]  \right\} 
\label{bosaction}  
\end{eqnarray}
where $\mathcal{ J}$ is a function appearing in the measure defining the operator $ {\mathcal P}$,
\be
h = h_0 -  \sum_K g\K   F_K^{\dagger} F_K  \label{hoperator}
\ee
\be
\nabla_t \, f = { 1 \over \tau} \left( f_{t+1} - f_t \right)  ,
\ee
\be
\Phi_t = {1 \over \sqrt{\Omega}} \sum_J (\beta_J)_t B_J^{\dagger} = {1 \over \sqrt{\Omega}}  \beta_t \cdot
B^{\dagger} 
\ee
\be
\Gamma_t= \left( 1\!\!1 + \Phi_t^* \Phi_{t-1}\right)^{-1}  
\ee
and $[..,..]_+$ is an anticommutator. Notice in the second line a trace inside the trace. The variables $\beta^*, 
\beta$  are always understood at times $t,t-1$ respectively. $S_{\mbox{eff}} $ has a global $U(1)$ symmetry
which implies boson conservation.

 The fermion interactions with external fields are  expressed in terms of the
bosonic terms which involve the matrix ${\cal M}$ (appearing in $h$). 

The dynamical problem of the interacting (composite) bosons can be
solved within the path integral formalism. 

 Part of the dynamical problem is the determination of the structure matrices $B_J$. This can be done by expressing the
energies in terms of the $B_J$ and applying a variational procedure which gives rise to an eigenvalue equation.

$S_{\mbox{eff}}$ must be compared to the action of elementary bosons. If $ H_B(b^{\dagger},b)$ is the
Hamiltonian of these bosons in normal form, the corresponding action is ~\cite{Nege}
\be
S_B = \tau \sum_t \left\{ \beta^* \nabla_t \beta - H (\beta^*,\beta) + \mu_B \beta^* \beta \right\}
\ee
where again the  variables $\beta^*, \beta$  are understood at times $t,t-1$ respectively.

We  notice that $ S_{\mbox{eff}}$ differs from $S_B$ in many respects

i) there is no canonical time derivative term 

ii) the coupling of the chemical potential ( appearing in $h$ is also noncanonical )

iii) there are non polynomial interactions because of the $\Gamma$-function. This function  becomes singular,
 as it will become clear in the sequel, when the number of bosons is of order $\Omega$, reflecting the Pauli
principle.

\section{The bosonic Hamiltonian}

Let us start by examining  the features of compositness when the number of bosons is much smaller than $\Omega$. Since
the expectation  value of $\beta^* \cdot \beta $ is of the order of the number of bosons, in this case we can perform an
expansion of logarithm and  $\Gamma$-function in inverse powers of $\Omega$. Expanding the logarithm we have
\begin{eqnarray}
& & { 1\over 2 \tau} \,  \mbox{tr} \ln \left[ 1\!\!1 + 
 \tau \Gamma \Phi^* \nabla \Phi \right] = 
{ 1\over 2} \mbox{tr} \left(\Phi^* \nabla_t \Phi \right)
\nonumber\\
& &  - { 1\over 4 } \mbox{tr}  \left[ \Phi^* \Phi \, \Phi^* \nabla_t\Phi \right] +... 
\end{eqnarray}
The first term can be made canonical by normalizing the boson form factors as the potential form factors
 \be
\mbox{tr} (B^{\dagger}_J \,  B_K ) = 2 \Omega \, \delta_{J,K}. \label{norm}
\ee
The other terms are then of order $\Omega^{-1}$.
Notice that the diagonal condition is only a matter of normalization, but the off diagonal one must be compatible
with the dynamics. If this is not the case, a redefinition of the $\beta$'s is necessary. 

Expanding the $\Gamma$-function we get the following couplings of the nucleon chemical potential
\be
\mu_N  \, \mbox{tr} \left[ \Phi_t^* \Phi_{t-1} - { 1 \over 2} ( \Phi_t^*  \Phi_{t-1})^2 +...\right]. \label{noncan}
\ee
Only the first term is canonical, and $\mu_N$ is not half the boson chemical potential as one might expect. As we will
see in Sec. V, for $ n \ll \Omega$ these anomalous couplings can be eliminated by a redefinition of the chemical
potential, so that in the case of a small number of bosons the Hamiltonian can be derived without difficulty.

But when the number of bosons is of order $\Omega$, the expansion of logarithm and $\Gamma$-function
 can be performed only after an appropriate subtraction, which can be performed by the change of variables
\be
\beta_K = \left( 1 -{1\over \Omega} r^2\right)^{ - { 1\over 2}}\,\, \beta'_K,  \label{transfor}
\ee
where $r$ is a parameter which will be fixed later. {\it This subtraction corresponds to the Bogoliubov
transformation in other approaches, but does not violate nucleon number conservation}.
We can then rewrite the $\Gamma$-function in the form
\be
\Gamma =  \left( 1 -{1 \over \Omega} \, r^2 \right) \Gamma',
\ee
where
\be
\Gamma'=
\left[ 1\!\!1 +  { 1 \over \Omega}\left(
\beta'^* \cdot B \,\,\beta' \cdot B^{\dagger} -  r^2 \right) \right]^{-1}.
\ee
For a suitable choice of $r$, $\Gamma'$ admits an expansion in $\Omega^{-1}$. Now we take the function $\mathcal{ J} $ 
appearing in $S_{\mbox{eff}}$ equal to the jacobian of the transformation ~(\ref{transfor})
\be
\mathcal{ J} = \left( 1 -{1\over \Omega} r^2\right)^{-N_B}  \label{jacobian}
\ee
where $N_B$ is the number of bosonic degrees of freedom.

Therefore the partition function becomes
\be
Z_C = \int \left[ {d \beta'^* d \beta' \over 2 \pi i} \right] \exp (- S')
\ee
where
 \begin{eqnarray}
& & S'(\beta'^*,\beta') = \tau \sum_t \mbox{tr} \left\{ { 1\over 2 \tau}   \ln  \left( 1\!\!1 + 
\tau  \, \Gamma' \, \Phi'^* \nabla_t \Phi'
 \right)  \right.
\nonumber\\
& & \left.  - { 1 \over 4}  \sum_K g_K  \left( 1 -{ r^2 \over \Omega} \right) \left[    \Gamma' \Phi'^*  F_K^{\dagger}
\, \mbox{tr} \, ( \Gamma' F_K \Phi' )  \right. \right.
\nonumber\\
& & \left. \left. +  2 \,  \left( \Gamma'  \,  F^{\dagger}_K F_K \right)
-  [ \Gamma' \Phi'^* F_K^{\dagger}, \Gamma'  F_K \Phi']_+   
  \right] \right.
\nonumber\\ 
& & \left.  + { 1 \over 2}   \left[ \Gamma'  \Phi'^*  
\, ( \Phi' \, h^T + h \,\, \Phi' ) \right]  + {1 \over 2} \sum_K g_K \left( F^{\dagger}_K
F_K \right) \right\}
\nonumber\\   
\end{eqnarray}
with
\be
\Phi'_t = {1\over \sqrt{\Omega}}  \beta'_t \cdot B^{\dagger}.
\ee
We assume, and we will verify a posteriori, that the parameter $r$ can be chosen in such a way that the anomalous time
derivative terms  be of order $ \Omega^{-1}$. Then to this order $Z_C$ can be written as a trace in a boson space 
\be
Z_C= \mbox{tr} \exp \left( - {1 \over T} H'\right).
\ee
The Hamiltonian $H'$ is obtained \cite{Nege} by omitting time derivative  and  chemical potential terms,
and replacing the variables $\beta'^*,\beta' $ by corresponding creation-annihilation operators
$b^{\dagger}, b$. These  satisfy canonical commutation relations and should not be confused with the
corresponding operators for the composites which are distinguished by a hat
\begin{eqnarray}
& & H'(r, \mu_N) = \,\, : \mbox{tr}\left\{ 
 - { 1\over 4}  \sum_K g_K  \left( 1 -{ r^2 \over \Omega} \right)  \left[   \Gamma_b \Phi_b^{\dagger}  F_K^{\dagger} 
 \right. \right.
\nonumber\\
& &  \left. \left. \times  \, \mbox{tr}(\Gamma_b F_K \Phi_b) +  2 \,  \Gamma_b  \,  F^{\dagger}_K F_K 
-  [ \Gamma_b \Phi_b^* F_K^{\dagger}, \Gamma_b  F_K \Phi_b]_+   
  \right] \right.
\nonumber\\ 
& & \left.  + { 1\over 2}  \left[ \Gamma_b  \Phi_b^*  
\, ( \Phi_b \, h^T + h \,\, \Phi_b ) \right]  + {1 \over 2} \sum_K g_K F^{\dagger}_K F_K
\right\}:. 
\nonumber\\   
\end{eqnarray}
The colons denote normal ordering and  
\begin{eqnarray}
\Phi_b & = & {1\over \sqrt{\Omega}} b \cdot B^{\dagger}
\nonumber\\
\Gamma_b &=&
\left[ 1\!\!1 + { 1 \over \Omega} \left(
b^{\dagger} \cdot B \,\,b \cdot B^{\dagger} - r^2\right) \right]^{-1}.  \label{gammab}
\end{eqnarray}
Here we meet with a subtlety. $H'$ commutes with the boson number operator, so we can select sectors with
a given number of bosons. But we are not guaranteed that these bosons carry nucleon number 2, because of
the noncanonical coupling of the chemical potential. We can enforce this fundamental property in the 
following way. Let us denote by $E'_0(n)$ the lowest eigenvalue of $H'$ in the sector of $n$ bosons. We  
require that $E'_0(n)$  be the lowest eigenvalue for
$n ={ 1 \over 2 }\, n_N$
\be
{\partial \over \partial n} E_0'| =0, \,\,\, n = { 1 \over 2}n_N. \label{r}
\ee
This determines $r$ as a function of the number of bosons and the nucleon chemical potential: $\overline{r}
=\overline{r}(n, \mu_N)$, ensuring that the bosons carry nucleon number 2.
 Condition (\ref{number}) then determines the nucleon chemical potential as a function of $n$: $
\overline{\mu}_N = \overline{\mu}_N(n)$. The boson Hamiltonian in the sector of n bosons is finally
\be
H_B(n)= H'(\overline{r}, \overline{\mu}_N) + 2 \, \overline{\mu}_N \,n.
\ee
It depends  on $n$ explicitly and through the dependence on $n$ of  $\overline{r},\overline{\mu}_N$. Therefore
also the matrices $B_J$ will depend on $n$, namely the form factors of the bosons depend on the number of the nucleons.

Notice that $H'$ provides the mapping of the nucleon interactions with external fields 
\be
c^{\dagger}{\cal M}c \rightarrow  : { 1\over 2}  \mbox{tr} \, \left[ \Gamma_b  \Phi_b^*  
\, ( \Phi_b \,{\cal M}^T + {\cal M} \,\, \Phi_b ) \right]: \,\,  .
\ee

\subsection{$s$-boson condensates}
 
This  procedure becomes particularly  simple if  the ground state contains only one species of bosons, 
which, having in mind the Interacting Boson Model, we call $s$-bosons. We will refer to such a ground state as to
an $s$-boson condensate". In such a case, assuming for simplicity
\be
F_0^{\dagger} F_0  =  B_0^{\dagger} B_0 = 1\!\!1,     \label{Bzero}
\ee
in the evaluation of the ground state energy we can set
\begin{eqnarray}
\Phi_b & = & {1\over \sqrt{\Omega}}  b_0 B_0^{\dagger}
\nonumber\\
\Gamma_b &=&
\left[ 1 + { 1 \over \Omega }\left(
b_0^{\dagger} b_0  -  r^2 \right) \right]^{-1}.  
\end{eqnarray}
It is then convenient to adopt the following definitions
\begin{eqnarray}
\hat{g}_K &=& { 1 \over 4 \Omega^2} |\mbox{tr}(B_0F_K^{\dagger})|^2 g_K
\nonumber\\ 
\hat{G} & = &\sum_K \hat{g}_K
\nonumber\\
G &=&\sum_K g_K.
\end{eqnarray}

Let us disregard for a moment normal ordering. This is equivalent to the semiclassical approximation in
the path integral formalism, namely to neglect quantum fluctuations. We then get for the lowest eigenvalue of
$H'$ in the $n$ boson sector
\begin{eqnarray}
& & E_0' = - \left( 1 + {1 \over \Omega}(n-r^2) \right)^{-1} \left(  \Omega G \, \left( 1- { r^2 \over \Omega}\right)
- 2 \overline{h} \, n \right) 
\nonumber\\
& & + \Omega G -\left( 1 + {1 \over \Omega}(n-r^2) \right)^{-2} \left( 1- { r^2 \over \Omega}\right) \,n 
\left( \Omega \hat{G} - G \right)
\nonumber\\
\end{eqnarray}
where
\be
\overline{h} = { 1 \over 2 \Omega}\mbox{tr} \, h.
\ee
Condition (\ref{r}) determines $r$ as a function of $n_N, \mu_N$
\be
\overline{r}^2 = { -2(\Omega +n) \overline{h} + \Omega (\Omega -2n) \hat{G}  -2 \Omega G
\over \Omega( -2 \overline{h} -2G + \Omega \hat{G}) }.
\ee
Inserting this value in $E_0'$ we get
\be
E_0' = -{\Omega \over \Omega \, \hat{G}- G } \left[ \overline{h} - { 1 \over 2} ( \Omega \hat{G} -2 G)  \right]^2.
\ee
Condition (\ref{number}) then  determines $\mu_N$
\be
\overline{\mu}_N=\overline{ e} - { 1 \over 2} \Omega \, \hat{G} + n \, \hat{G}  - { n \over \Omega} \, G
\ee
where
\be
\overline{e}= { 1 \over 2 \Omega}\mbox{tr} \, e.
\ee
It is not surprising that with this value of $\overline{ \mu}_N$
\be
\overline{r}^2 = n.
\ee
We then get also the lowest eigenvalue of $H_B$ (neglecting normal ordering) 
\be
E_0 = E_0' + 2 \mu_N n = 2 n \, \overline{ e } - n \, \Omega \, \hat{G} + n^2 \left( \hat{G} - { 1 \over \Omega} \, G
\right).
\ee
In the case of a monopole pairing interaction, $ \hat{ G} = G = g_0$, comparing to the exact spectrum (\ref{pairspec}),
we see that  the coefficients of the powers of $n$ are affected by errors  of order $\Omega^{-1}$. The form factor of
the $s$-boson is determined by the minimizing $E_0$
\be
B_0 = F_0.
\ee
In order to get an expression of $H_B$ of practical use, we must perform an expansion in $\Omega^{-1}$. Since 
the energy scale is set by the single-particle energies, we must make an assumption concerning the
magnitude of the coupling constants $g_K$ with respect to $\overline{e}$. For a system with infinitely many 
degrees of freedom, $ \Omega \rightarrow \infty $, in order to get finite energies we must assume $ g_K \sim
\Omega^{-1}$, in which case $\mu_N $ is of order $\Omega^{0}$. Such a behavior is also acceptable for many nuclei.

To take into account quantum effects we must put $H'$ in normal order. This corresponds to include quantum  
fluctuations in the path integral formalism, and requires an expansion with respect to $\Omega^{-1}$. Remaining
in the case in which the ground state is an $s$-boson condensate, we only need the following equation
\begin{eqnarray}
 & & { 1 \over n!}\< b_0^n \,  \sum_{s=0}^{\infty}  c_s : \left(b_0^{\dagger} b_0 \right)^s :  \, \left(
b_0^{\dagger}\right)^n \> 
\nonumber\\
 & & =
\sum_{s=0}^n c_s  n ( n-1) ...(n-s+1) .
\end{eqnarray}
Notice that the expectation value of any normal ordered function in a state of $n$ bosons is a polynomial of degree not
greater than $n$.

A further simplification occurs if the number of the other bosons, which we denote by the label $\overline{K}$,
is much smaller than $\Omega$. In such a case  we can obviously classify the terms appearing
in the function $ \Gamma_b$, Eq.~(\ref{gammab}), according to
\begin{eqnarray}
& &   b_0^{\dagger} \,b_0  - r^2 + \sum_{\overline{K}_1,{\overline{K}_2}}
b_{\overline{K}_1}^{\dagger}  b_{\overline{K}_2} \,  B_{\overline{K}_1} \, B^{\dagger}_{\overline{K}_2} 
\,\, \sim \,\,\, 1
\nonumber\\
& & b_0^{\dagger} \,  B_0 \sum_{\overline{K}}
 b_{\overline{K}}  \, B^{\dagger}_{\overline{K}} \,\,\, \sim \,\,\,\sqrt{\Omega}.
\end{eqnarray}
It is then easy to see that neglecting terms of order $ \Omega^{-{ 1 \over 2}}$ or smaller, $H_B$ is at most
quartic in the $\overline{K}$ boson operators.

\section{An alternative derivation of the Hamiltonian for $ n \ll \Omega$}

In this Section we restrict ourselves to the case of a small number of bosons. Then the subtraction is
not necessary, we can set $ \overline{r} = 0 $, and we can put the effective
action in canonical form by a shift of the chemical potential. It is to emphasized that no other quantum
corrections are necessary after such a shift. 
 
 For simplicity we assume the coupling  $g_0$ to be positive (attractive pairing force)  and larger than the other ones,
so that at the minimum  only $\beta_0$ is different from zero. Assuming $B_0$ to satisfy Eq.(\ref{Bzero}), 
$\mbox{S}_{\mbox{eff}}$ at constant fields is
\begin{eqnarray}
T {\overline S} & =& \Omega  \sum_K g\K -  \Omega \left( g_0 |\beta_0|^2 + \sum_K g\K \right) 
\left( 1 + { 1\over \Omega} |\beta_0|^2 \right)^{-2} 
\nonumber\\
& & + { 1 \over \Omega} \mbox{tr} \, (e-\mu_N) \, |\beta_0|^2 \left( 1 + { 1\over \Omega} |\beta_0|^2 \right)^{-1}. 
\end{eqnarray}
Its  minimum with respect to $ |\beta_0|^2 $ can be determined exactly, but since we will perform the $ 1 / \Omega$
expansion we put ourselves in this framework since the beginning. We will retain only the first order corrections, which
are of order $\Omega^0$, with the exception of the coupling with external fields which are of order $\Omega^{-1}$.
We remind that the  difficulties in the derivation of the boson Hamiltonian are due to anomalous time derivative
terms and couplings of the chemical potential. In this approximation the first difficulty is overcome because, as
already noted, noncanonical time derivatives are of order $1 / \Omega$. It remains to get rid of the noncanonical
couplings of the chemical potential. For this purpose we set
\be
\mu_N = \Omega \mu_1 + { 1 \over 2}  \mu_B \label{mu}
\ee
and expand wr to $ 1/ \Omega$
\be
T {\overline S}  =  - \Omega ( 2 \mu_1 + g_0 ) | \beta_0|^2 +  (2  \overline{e} - \mu_B )| \beta_0|^2
+ 2 ( \mu_1 + g_0 ) | \beta_0|^4.
\ee 
Since $| \beta_0|^2 \ll \Omega$, at the minimum the first term must vanish separately from the others and we get 
\begin{eqnarray}
\mu_1 &=& - { 1 \over 2} g_0
\nonumber\\
| \beta_0|^2 & = & { 1 \over g_0}( { 1 \over 2} \mu_B - \overline{e}).
\end{eqnarray}
We select  a sector with a given number $n $ of bosons by imposing the condition
\be
{ \partial \over \partial \mu_B} ( T {\overline S}) =  n
\ee 
which yields
\be
|\beta_0|^2 = n .
\ee
We see from Eq.~(\ref{noncan}) that to order $ \Omega^0$ the only noncanonical term is proportional
to $\mu_1$, which does not depend on $n$. We can then  insert in the action the definition~(\ref{mu}) and get  a
canonical bosonic action with canonical chemical potential $\mu_B$. 

There remains a last point. The energies are given by (minus) the logarithm of the partition function plus
the chemical potential times the number of bosons. But we can subtract from the action the term
$\Omega\mu_1\beta^*_t  \cdot \beta_{t-1}$, and subtract in the end from the energy only $\mu_B$ 
times the number of bosons. In this way we get exactly the boson Hamiltonian $H_B$.

\subsection{Few bosons in a single $j$-shell}

If the nucleons live in a single $j$-shell the form factors of the composites either vanish, or are equal to the  form
factors of the potential. But unless $I << j$  condition (\ref{cond}) is not satisfied, the composites
do not have a high index of nilpotency and must be excluded.

 We identify the quantum number $K$ with the boson angular momentum, $ K = (I_K,M_K)$, so that the form factors of the
potential are proportional to Clebsh-Gordan coefficients
\be
(F_{IM})_{m_1,m_2}= \sqrt{2 \Omega} \, \<j m_1 j m_2| IM \>, \,\,\, \Omega = j + { 1\over 2},   
\ee
with the conventions of~\cite{Vars}.

The resulting action is
\begin{eqnarray}
& & S(\beta^*,\beta)  = \sum_t \left\{ \sum_{I_1 I_2} \beta^*_{I_1} \left[ ( \nabla_t -  \mu_B )  +
\omega \right]_{I_1 I_2}  \beta_{I_2} 
\right.
\nonumber\\ 
 & &  \left.  +  \sum_{I_1 I_2 I_3 I_4}  \sum_{IM}  W^I_{I_1 I_2 I_3 I_4} \left( \beta^*_{I_1} \,
\beta^*_{I_2}\right)_{IM}\left( \beta_{I_3} \, \beta_{I_4}\right)_{IM} \right\}
\nonumber\\
\end{eqnarray}
where also for the $9j$ symbol we adopt the conventions of \cite{Vars}, and
\begin{eqnarray}
\omega_{I_1 I_2} &=& {1 \over \Omega} \mbox{tr} \left(\, F_{I_1} F^{\dagger}_{I_2} \, e \right)
 -g_{I_1} \Omega \, \delta_{I_1 I_2}
\nonumber\\
 W^I_{I_1I_2I_3I_4} &=&  \left( - 2 g_0  + \sum_{i=1}^4 g_{I_i}  \right)  \Pi_{i=1}^4 [(2I_i+1)]^{1/2}  
\nonumber\\
& & \times  \Omega 
  \left\{
    \begin{array}{ccc}
       j & j & I_1   \\
       j & j & I_2   \\
       I_3 & I_4 & I \\
    \end{array}
     \right\} 
\nonumber\\
\left( \beta_{I_3} \,  \beta_{I_4}\right)_{IM} & = &\sum_{M_3 ,M_4} \<I_3,M_3,I_4,M_4|I,M\> \, 
\beta_{I_3 M_3} \beta_{I_4 M_4}.
\nonumber\\
\end{eqnarray}

The Hamiltonian is obtained \cite{Nege} by omitting the time derivative  and  chemical potential terms,
and replacing the variables $\beta^*,\beta $ by corresponding creation-annihilation operators
$b^{\dagger}, b$. These  satisfy canonical commutation relations and should not be confused with the
corresponding operators for the composites which are distinguished by a hat
\begin{eqnarray}
 H_B &=& \sum_{I_1M_1I_2M_2} \omega_{I_1M_1I_2M_2} b^{\dagger}_{I_1M_1}  b_{I_2M_2}  + 
\sum_{I_1I_2I_3I_4} 
\sum_{IM} 
\nonumber\\ 
 & &  \left\{ W^I_{I_1I_2I_3I_4} \left( b^{\dagger}_{I_1} \,
b^{\dagger}_{I_2}\right)_{IM}\left( b_{I_3} \, b_{I_4}\right)_{IM} \right\}.   \label{bosHam}
\end{eqnarray}
It is easy to check that, due to the symmetries of the 9j symbols, $H_B$ is hermitian.

From the interaction with external fields we get the fermion-boson mapping of other operators
\begin{eqnarray}
 c^{\dagger} {\cal M} c & & \rightarrow  \sum_{I_1M_1I_2M_2}  { 2 \over  \Omega}
\mbox{tr} \left( F_{I_1M_1} {\cal M} \, F_{I_2M_2}^{\dagger} \right)  b^{\dagger}_{I_1M_1} b_{I_2M_2} 
\nonumber\\
& &
 + \sum_{\mbox{all} \, I,M} \left({ 2 \over  \Omega}\right)^2 
\mbox{tr} \left( F_{I_1M_1} {\cal M} \, F_{I_4M_4}^{\dagger} F_{I_2M_2} F_{I_3M_3}^{\dagger}\right) 
\nonumber\\
& & 
 \times b^{\dagger}_{I_1M_1} b^{\dagger}_{I_2M_2} b_{I_3M_3} b_{I_4M_4}.
\end{eqnarray}
 Since the above Hamiltonian has been derived under the restriction $n<< \Omega$ in a single
$j$-shell, we can assume 
\be
e_{m_1 m_2} = \overline{e} \, \delta_{m_1 m_2},
\ee
so that the single boson energy matrix is diagonal
\be
\omega_{I_1 I_2} = ( 2 \, \overline{e} - g_{I_1} \Omega) \delta_{I_1 I_2}.
\ee
 But the bosonic interactions couple all the bosons with angular momenta for which the $9j$ symbols do not vanish,
even if the corresponding potentials do vanish. It is however important to remember that the
coefficients of the terms involving high angular momentum can change after the dependence of the index of nilpotency
on the angular momentum is taken into account.

\subsection{ Monopole  pairing }

Let us consider the case of a pure monopole pairing interaction, namely $g_I=0, I\neq 0$. The exact
spectrum is
\be
E_{n,v} = 2 n \overline{e} - g_0 ( \Omega +1) (n-v) + g_0 (n-v)^2 + g_0 (n-v) v.  \label{pairspec}
\ee
We adopted definitions slightly different from the usual ones but more convenient, we think, in
the present context.  Here $n$ is the total number of pairs, ie half the number of nucleons, and $v$ is half the
standard seniority, ie the number of pairs not coupled to zero angular momentum. It is then natural to relate the number
of $s$-bosons $n_s$ to $n$ and $v$ according to
\be
n_s = n-v . 
\ee
In terms of $n_s,v$, the eigenvalues can be written
  \be
E_{n_s +v,v} = 2 n \overline{e} - g_0 ( \Omega +1) n_s + g_0 n_s^2 + g_0 n_s (n-n_s) .
\ee
Setting $b_0= s$ and denoting by $\overline{K}$ angular momenta greater than zero
we get from~Eq.(\ref{bosHam}) the bosonized pairing Hamiltonian
\begin{eqnarray}
& & H_{\mbox{pairing}} = \left[ 2 \overline{e} - g_0 (\Omega+1) \right] \, s^{\dagger} \, s  + g_0 s^{\dagger}\,
\, s \,s^{\dagger} s
\nonumber\\
& & + g_0 \sum_{\overline{K}_1,\overline{K}_2,\overline{K}_3}  ( 2 \overline{K}_1+1)^{-{ 1\over 2}}
 [(2 \overline{K}_2 +1)(2 \overline{K}_3 +1)]^{{ 1\over 2}}  
\nonumber\\
& & \times   <\overline{K}_20\overline{K}_30|\overline{K}_10>\left[ (s^{\dagger}
b^{\dagger}_{\overline{K}_1})_{\overline{K}_1}
\cdot (b_{\overline{K}_2}b_{\overline{K}_3})_{\overline{K}_1} + h.c.
\right]
\nonumber\\
  & & + \sum_{IM} v_2^I (d^{\dagger} \, d^{\dagger})_{I M}(d \, d)_{I M}.
\end{eqnarray}
The expression of $v^I_2$ is given in the next Section.
The sector of zero seniority, $n=n_s$, decouples and has the exact spectrum. The study of the seniority spectrum, in
which the dependence of the index of nilpotency on the angular momentum might be important, is left for a
future work.

\subsection{Monopole plus quadrupole pairing}

We consider now the case in which the nuclear Hamiltonian contains  a pairing plus quadrupole pairing
interaction. One might hope in a repetition of the pattern found in the  bosonization of the pairing model, namely that 
the $s-d$ bosons should decouple from the others, but this does not happen. We can write the Hamiltonian in the form
\be
H = H_{s-d} + H_Q,
\ee
where $H_Q$ contains at least one boson with angular momentum greater than 2 and $ H_{s-d} $ is the Hamiltonian
truncated to the $s-d$ subspace  
\begin{eqnarray}
H_{s,d} &=& \omega_0 \, s^{\dagger} \, s + \omega_2  \, d^{\dagger} \,\cdot  d 
+ v_0 s^{\dagger}\, s^{\dagger} \, s \, s + 
 w \, s^{\dagger} \, s  \,\, d^{\dagger} \cdot d +
 \nonumber\\ 
 & & \sum_I v_2^I (d^{\dagger} \, d^{\dagger})_I \cdot (d \, d)_I
+ v_{02} \left[(d^{\dagger} d^{\dagger})_{0,0} s \, s + h.c.\right]
\nonumber\\
 & &+ \tilde{v}_{0,2} \left[ \left( d^{\dagger}  \, s^{\dagger}\right)_2 \cdot 
\left(d \, d \right)_2 + h.c. \right]
\end{eqnarray}
The parameters in the above equation are
\begin{eqnarray}
\omega_0 &=&  2 \, \overline{e} - g_0 \Omega
\nonumber\\
\omega_2 & =& 2 \, \overline{e} -g_2 \Omega
\nonumber\\
v_0 &=& 2 \Omega 
 \left\{
    \begin{array}{ccc}
       j & j & 0   \\
       j & j & 0   \\
       0 & 0 & 0 \\
    \end{array}
     \right\} g_0 = g_0
\nonumber\\
v_2^I &=& 50 \Omega 
\left\{
    \begin{array}{ccc}
       j & j & 2  \\
       j & j & 2   \\
       2 & 2 & I \\
    \end{array}
     \right\} ( 2 g_2 - g_0)
\nonumber\\
w & =& 40 \Omega
\left\{
    \begin{array}{ccc}
       j & j & 0 \\
       j & j & 2   \\
       0 & 2 & 2 \\
    \end{array}
     \right\} g_2 = 4 g_2 
\nonumber\\
v_{0 2} &=& 10  \Omega
\left\{
    \begin{array}{ccc}
       j & j & 2   \\
       j & j & 2   \\
       0 & 0 & 0 \\
    \end{array}
     \right\}  g_2 = \sqrt{5} g_2
\nonumber\\
\tilde{v}_{02} &=& 
 10 \sqrt{5} \Omega
\left\{
    \begin{array}{ccc}
       j & j & 2  \\
       j & j & 0   \\
       2 & 2 & 2 \\
    \end{array}
     \right\} (3 g_2 -g_0 ) 
\nonumber\\
 & \sim & \sqrt{{ 10 \over 7 } } ( 3 g_2 - g_0 ).
\end{eqnarray}
The last approximate equality holds for $j \gg1$.
Now in order to see if we have an at least approximate decoupling of the $s-d$ sector, we might proceed in the
following way.
After introducing the dependence of the index of nilpotency on the angular momentum, we should evaluate the energy of
the $s$- and $d$-bosons first by assuming that the form factors of all the others vanish, then assuming nonvanishing
only the form factor of the boson with angular momentum 4 and so on. If these energies get their minimum when all the
other form factors vanish, we have an exact decoupling. Otherwise we can have an approximate decoupling if these
energies depend little on the inclusion of higher momentum bosons. We leave this investigation for a future work
\cite{Loiu} in which the determination of the form factors of the composites will be studied in a nucleon space
including many shells.

\section{The path integral formalism: The Goldstone and Higgs fields}

In this Section we outline the treatment of the effective action in the standard path integral
formalism. This can be helpful in a numerical simulation.

We must first determine the  classical value of the $s$-boson field without breaking nucleon conservation. This
can be done adopting the polar representation for this field
\be
\beta_0= r \exp \left( i \, { \theta \over \overline{r}} \right)
\ee
\be
r^2 = \overline{r}^2 \left( 1 + { \sigma \over \overline{r}} \right).
\ee
Boson number conservation is a consequence of the invariance of the action under the transformation
\be
\beta_0 \rightarrow \beta_0 \exp(i \alpha).
\ee
It leaves $r$, and therefore $\sigma$, invariant, while
\be
\theta \rightarrow \theta + \alpha.
\ee
We call~\cite{Barb} $\theta$ and $r$ the Goldstone and Higgs fields by analogy with the Godstone model.
$\overline{r}$ is the classical field. The integration over $\sigma$ in the partition function extends from $ -
\overline{r}$ to $\infty $, and one has to devise different approximations depending on the value of $  \overline{r}$.
All the other terms in the action must be expressed in terms of the Goldstone and Higgs fields. For the time derivative
terms, for instance, we have
\begin{eqnarray}
\sum_t \beta^*  \cdot \nabla_t  \beta &= & \sum_t \left\{ - { 1\over 8} \tau^2 \left( \nabla_t^{(-)} \sigma \right)^2
 - { 1\over 2} \tau^2 \left( \nabla_t^{(-)} \theta \right)^2 \right.
\nonumber\\
& & \left. -i \tau \sigma \nabla_t^{(s)} \theta + \sum_{\overline{K}} \beta^*_{ \overline{K}}  \cdot \nabla_t \, 
\beta_{
\overline{K}} 
\right\}
\end{eqnarray}
where $ \overline{K}$ refers to all the other bosons and $\nabla_t^{(s)} $ is the symmetric time
derivative. The path integral must then be evaluated as a function of $\overline{r}$ and $ \mu_N$, and these parameters
are fixed by minimizing the free energy and by imposing condition~(\ref{number}) respectively. The evaluation 
of the path integral is performed by first determining the minimum of the action at constant fields, and then
including the contribution of quantum fluctuations. But at the end we must check that $2n = n_N$.

\section{Summary and outlook}

 We have presented a  new method of bosonization in which we restrict the partition function of the nucleus to
nucleon composites. We get in this way the euclidean action of the effective bosons in closed form respecting all
nucleon symmetries, in particular nucleon number conservation. Indeed the presence of a large number of nucleons
in the ground state is accounted for by a subtraction which does not violate nucleon number conservation. The only
approximation made concerns the replacement of the identity operator in the space of the composites by 
the operator ${\mathcal P}$. 

 The nuclear dynamics  can  be studied by the methods of path integrals, including numerical simulations which now are
not affected by the sign problem, or in the more usual Hamiltonian formalism. Since the effective
boson action does not have a canonical form, it has been necessary to devise an appropriate procedure to derive the
boson Hamiltonian. This can be put in a form useful for practical applications, however, only by performing an
expansion in the inverse of the index of nilpotency of the composites.

The formalism is consistent only if all the composite bosons involved have a high index of nilpotency. Obviously bosons
with high angular momentum do not satisfy such condition, as shown by the example of a boson with maximum
angular momentum
 \be
\hat{b}_{2j-1,2j-1} = \sum_{m_1 m_2} \<jm_1 jm_2|2j-1,2j-1\> c_{m_1} c_{m_2} \label{nonbos}
\ee
which has index of nilpotency 1.
It is then necessary to ascertain that bosons with low index of nilpotency
will decouple. This problem sometimes is not explicit in some  approaches, because the boson space is truncated
directly or as a consequence of a truncation of the nucleon space. In this connection it is important to observe that
the restriction to a definite boson space, like in the Interacting Boson Model, does not require an exact
decoupling. An approximate decoupling is sufficient, because states weakly coupled can  easily be
integrated out in the path integral before deriving the Hamiltonian. This fundamental feature is left for a future
investigation.

Another important point is the inclusion of particle-hole terms in the nucleon-nucleon potential. This requires the
introduction of fonons, which will be done by an appropriate extension of the present technique.

It is perhaps worth while to emphasize that for a derivation of the Interacting Boson Model it is not at all
necessary that phonons be unimportant. It is sufficient that they can be integrated out like pions in the
derivation of the nucleon-nucleon potential.

In a future paper we will discuss in detail the determination of the form factors
of the composites~\cite{Loiu}. This can be done by adopting the natural parametrization
\be
(B_{J,M})_{m_1,m_2}= \sum_{j_1 j_2} p^J_{j_1 j_2} \<j_1 m_1 j_2 m_2|JM\>.
\ee
The energies of the bosons are functions of the matrices $p^J$ which can be determined by a  variational 
calculation.

\begin{acknowledgments}
I am grateful to N. Loiudice for many fruitful discussions.
\end{acknowledgments}

\appendix
\section{Basic formulae for Berezin integrals \label{A}}

In this Section we report for the convenience of the reader some basic formulae for Berezin integrals that
we need. Their definition for a single Grassmann variable is
\be
\int d \gamma ( a \gamma + b) = a,
\ee
the generalization to many variables being obvious. For a change of variables
\be
\gamma_i = \gamma_i ( \gamma')
\ee
in a multiple integral we have
\be
\int  \prod_i (d \gamma_i) f ( \gamma) = \left( \det {\partial \gamma_h \over \partial \gamma_k'} \right)^{-1}
\int \prod_i (d \gamma_i') f( \gamma').
\ee
Notice the appearance of the inverse of the jacobian, contrary to the case of ordinary variables.

Gaussian integrals can be evaluated exactly, like for ordinary variables. There are two types of such integrals
\begin{eqnarray}
\int \prod_h (d \gamma_h^* d \gamma_h) \exp \sum_{ij} \gamma_i^* M_{ij} \gamma_j &=& \det M
\\
\int \prod_h (d \gamma_h) \exp \sum_{ij} { 1 \over 2} \gamma_i A_{ij} \gamma_j &=& \mbox{Pf} \,A  \label{pfaffian}
\end{eqnarray}
where $\mbox{Pf} \,A$ is called the $pfaffian$ of $A$. The following algebraic identity holds
\be
\left( \mbox{Pf} \,A \right)^2 = \det A.
\ee

\section{Inner products of composite states \label{B} }

Let us first consider the case of only one composite. To evaluate the inner product of coherent states
we use the identity operator in the fermion Fock space
\be
{\cal I} = \int d \gamma^* d \gamma \<\gamma | \gamma\>^{-1} | \gamma \> \<  \gamma |
\ee
where the $\gamma^*,\gamma$ are Grassmann variables and $|\gamma \>$ coherent nucleon states
\be
|\gamma \> = \exp (- \gamma \, c^{\dagger}) \>.
\ee
We then have
\be
\<\beta_1|\beta\> = \<\beta_1| {\cal I} |\beta\>= \int d \gamma^* d \gamma \, \exp ( - \gamma^*  \gamma ) 
\<\beta_1 | \gamma\> \< \gamma |\beta\>.
\ee 
The matrix element $ \<\beta_1| \gamma\>$ can be evaluated  using the defining property of coherent states
\be
c | \gamma\> = \gamma| \gamma\>
\ee
with the result
\be
\<\beta_1|\gamma\> = \exp \left( { 1\over 2 \sqrt{\Omega}} \beta_1^* \gamma B \gamma \right).
\ee
Therefore $\<\beta_1|\beta\> $ becomes 
\be
\<\beta_1|\beta\> = \int d \gamma^* d \gamma \, E(\gamma^*,\gamma,\beta_1^*,\beta),
\ee
where the function $E$ is 
\begin{eqnarray}
 E(\gamma^*,\gamma,\beta^*,\beta)   & = & \exp \left( -\gamma^* \gamma + 
{1\over 2 \sqrt{\Omega}} \beta^* \, \gamma \, B \, \gamma \right.
\nonumber\\
 & & \left. +  { 1\over 2\sqrt{\Omega}} \beta  \, \gamma^* B^{\dagger} \gamma^* \right). \label{E}
\end{eqnarray}
 By the change of variables 
\be
\gamma' =  \gamma^* - { \sqrt{\Omega} \over \beta} (B^{\dagger})^{-1} \gamma
\ee
the integral is factorized according to
\begin{eqnarray}
& & \<\beta_1| \beta\> = \int \, d \gamma' \exp \left({ 1 \over 2 \sqrt{\Omega}} \gamma' \,  \beta\,
B^{\dagger}
\,
\gamma'
\right) 
\nonumber\\ 
 & & \,\,\, \times \int \, d \gamma \, \exp \left[ { 1\over 2} \gamma \left( \sqrt{\Omega} \left( \beta \,
B^{\dagger}\right)^{-1} + { 1 \over \sqrt{\Omega}}\beta_1^* \, B \right) \, \gamma \right].
\end{eqnarray}
The factors are of the form~\ref{pfaffian}, so that finally
\begin{eqnarray}
\<\beta_1|\beta\> &=&  \left[ \det \left( { 1 \over \sqrt{\Omega}}\beta B^{\dagger}\right) \right]^{{ 1 \over 2}}  
\left[ \det \left(   \sqrt{\Omega}  ( \beta B^{\dagger} )^{-1} + \right. \right.
\nonumber\\
& &
  \left. \left.{ 1 \over \sqrt{\Omega}}\beta_1^* B \right) \right]^{{1 \over 2}}\det \left[ 1\!\!1 + { 1 \over
\sqrt{\Omega}}\beta \, \beta_1^* B^{\dagger} B \right] ^{{1 \over 2}}.
\end{eqnarray}
 It is perhaps worth while noticing that in the limit of infinite $\Omega$, assuming the structure function to
satisfy the condition~(\ref{Bzero}) we recover exactly the expressions valid for elementary bosons, in particular
\be
\<\beta_t|\beta\>= \left(1+ {1\over \Omega} \beta_t^* \beta \right) ^{\Omega} \rightarrow \exp (\beta_t^*
\beta),\,\,\,
\Omega \rightarrow \infty.
\ee
We will further need the inner product
\be
\<\beta|( {\hat b}^{\dagger})^n \> = C_n (\beta^*)^n,
\ee
where
\be
C_n = { \Omega! \over (\Omega-n)! \Omega^n}= 
\left( 1- { 1\over \Omega}\right) \left( 1- { 2\over \Omega}\right)...
\left( 1- { n-1\over  \Omega}\right). 
\ee
In the general case of many composites the above equations become
\be
\<\beta_t|\beta\> = \left[ \det \left( 1\!\!1 + \Phi_1 \Phi \right) \right]^{1 \over 2},
\ee
\begin{eqnarray}
& &\<\beta_t| ({\hat b_{I_0}}^{\dagger})^{n_0}...{\hat b_{I_i}}^{\dagger})^{n_i}\> =
{\partial^{n_0} \over \partial x_0^{n_0} }... {\partial^{n_i} \over \partial x_i^{n_i} }
\nonumber\\
& &  \,\,\,\,\, \times \exp \left\{{ 1\over 2} \mbox{tr} \ln [ 1\!\!1 +{ 1 \over \sqrt{\Omega}}(x \cdot B^{\dagger})
\Phi^*_t]
\right\}|_{x=0}
\end{eqnarray}
where 
\be
  x \cdot B^{\dagger} =  \sum_J x_J 
B_J^{\dagger}.
\ee

\section{The operator $\mathcal{P}$ \label{C}}

In order to implement the assumption of Boson Dominance we need the projection operator in the space of the composites
describing the physical degrees of freedom which dominate the partition function. Guided by the comparison with
elementary bosons we  approximate this operator by $\mathcal{P}$. This comparison suggests that the measure
in $\mathcal{P} $ should be $\< \beta| \beta \>^{-1}$. But, not surprisingly, we find that the measure must
depend on the number of nucleons, namely of bosons 
\be
d \mu(\beta^*, \beta) = J^{-1} \< \beta| \beta \>^{-1},
\ee
where the Jacobian is given in Eq.~(\ref{jacobian}).

For this reason it is necessary to introduce two types of
composite bosons. We distinguish by a "bar" the new boson operators and their structure matrices. These new
bosons are to be regarded as the "physical" ones, while those appearing in $\mathcal{P}$ are auxiliary operators.

 Let us see the action of $\mathcal{P}$ on physical
states. Let us first consider the case in which there is only one physical composite with structure function satisfying
the equation
\be
\overline{B}^{\dagger} \overline{B} =  \overline{\gamma}^2 \, 1\!\!1.  
\ee
Using the results of Subsection B we find
\be
{\mathcal P}|(\overline{b}^{\dagger})^n \> = J^{-1} \left(\Omega \, \overline{\gamma}^2 \right)^{{n \over 2}} 
{ \Omega^2 \over ( \Omega - n)( \Omega - n-1)}|( \overline{b} )^n \>.
\ee
Setting $r^2 = n$,  ${\mathcal P} $ is the identity on states of $n$ bosons provided
\be
\overline{\gamma}^2 = { 1 \over \Omega} \left( { \Omega - n +1 \over \Omega } \right)^{ {2 \over n}}.
\ee
In the case of many composites, using the inequality
\be
\mbox{tr} \left( {1 \over\Omega} B^{\dagger} B \right)^n << \left( \mbox{tr}  \left({1 \over\Omega} B^{\dagger} B
\right) \right)^n
\ee
which follows from the condition~(\ref{cond}) 
we find again that ${\cal P}$ approximates the identity with an error of order
$ 1 /
\Omega$
\be
{\mathcal P}| (\overline{b}_{I_0}^{\dagger})^{n_0}...( \overline{b}_{I_i}^{\dagger})^{n_i}\> =
|  (\overline{b}_{I_0}^{\dagger})^{n_0}...(\overline{ b}_{I_i}^{\dagger})^{n_i} \, + O(1 / \Omega) \>.
\ee

\section{Derivation of the effective action \label{D}}

 For the following manipulations we need the Hamiltonian in antinormal form
\be
H = { 1\over 2} \mbox{tr} ( h + h_0 ) - c \,  h^T c^{\dagger} -
\sum_K g\K \,  {1\over 2} \, c F_K c
\, { 1\over 2} \, c^{\dagger} F^{\dagger}_K c^{\dagger}
\ee
where the upper script $T$  means "transposed" and $h$ was given in Eq.(\ref{hoperator}).
Now we must evaluate the matrix element $\< \beta_t| \exp(- \tau  H) |\beta_{t-1} \>$. To this end we expand to 
first order in
$\tau$ (which does not give any error in the final $\tau \rightarrow 0$ limit) and
insert the operator ${\cal P}$ between annihilation and creation operators
\begin{eqnarray}
& &\< \beta_t| \exp ( - \tau H ) |\beta_{t-1}\> = \exp \left( - {1\over 2} \mbox{tr}(h+h_0) \tau \right) \,\,\<\beta_t|
{\cal{P}}
\nonumber\\
& & \,\,\,\,\, 
 -  c \,  h^T \tau \, {\cal{P}} c^{\dagger}
 \sum_k g_k \tau  \, {1\over 2} \, c F_K c \, {\cal{P}}
\, { 1\over 2} \, c^{\dagger} F^{\dagger}_K c^{\dagger} |\beta_{t-1}\>.
\end{eqnarray} 
Using the identity in the nucleon Fock space we find
\begin{eqnarray}
& &\<\beta_t|  \exp (- \tau H ) |\beta_{t-1}\> =  \int d \gamma^* d \gamma  \, 
E(\gamma^*,\gamma,\beta_t^*,\beta_{t-1})
\nonumber\\
& & \,\,\,\,\,\,\,\,\,\, \times   \exp\left( - {1\over 2} \mbox{tr}(h+h_0) \tau   - \gamma^* h \, \tau \gamma \right)
\nonumber\\ 
& & \,\,\,\,\,\, \,\,\,\,\times 
\exp \left(   \sum_K g_K \tau \, {1\over 2} \gamma \,F_K \,\gamma  
 \, { 1\over 2}  \gamma^* F_K^{\dagger} \gamma^* \right)
\end{eqnarray}
where the function $E(\gamma^*,\gamma,\beta^*,\beta)$ is defined in~(\ref{E}).
By means of the Hubbard-Stratonovich transformation we can make the exponents quadratic in the Grassmann
variables and evaluate the Berezin integral
\begin{eqnarray}
& & \<\beta_t| \exp(- \tau  H) |\beta_{t-1} \> =  \int \prod_K da_K^* da_K \exp (   - a^* \cdot a  )
\nonumber\\
& & \,\,\, \times \exp \left\{ 
 { 1\over 2} \mbox{tr} \ln \left[ 1\!\!1 + \left( \Phi_t^* + \sum_{K_1} \sqrt{ g_{K_1}
\tau} 
\,
 a^*_{K_1} F_{K_1}\right)
\right. \right.
\nonumber\\
& & \,\,\, \left. \left.  \times R^{-1} \left( \Phi_{t-1} + \sum_{K_2} \sqrt{ g_{K_2} \tau} 
\, a_{K_2} (F_{K_2})^{\dagger}\right) (R^T)^{-1} \right] \right\}
\nonumber\\
& & \,\,\, \times  \det R \, \exp \left( - {1\over 2} \mbox{tr}(h+h_0) \tau \right),
\end{eqnarray}
where
\be
R = 1\!\!1 + h \, \tau.
\ee
Performing the integral over the auxiliary fields $a\K^*,a\K$ we get 
\begin{eqnarray}
& & \<\beta_t| \exp(- \tau  H) |\beta_{t-1} \> =  \int \prod_K da_K^* da_K \exp (   - a^* \cdot a  )
\nonumber\\
& & \,\,\, \times \exp \left\{ 
 { 1\over 2} \mbox{tr} \ln \left[ 1\!\!1 + \left( \Phi_t^* + \sum_{K_1} \sqrt{ g_{K_1}
\tau} 
\,
 a^*_{K_1} F_{K_1}\right)
\right. \right.
\nonumber\\
& & \,\,\, \left. \left.  \times R^{-1} \left( \Phi_{t-1} + \sum_{K_2} \sqrt{ g_{K_2} \tau} 
\, a_{K_2} (F_{K_2})^{\dagger}\right) (R^T)^{-1} \right] \right\}
\nonumber\\
& & \,\,\, \times  \det R \, \exp \left( - {1\over 2} \mbox{tr}(h+h_0) \tau \right),
\end{eqnarray}
The functional form of the composites partition function is
\be
Z_C = 
 \int \left[{d \beta^* d \beta \over 2 \pi i} \right]\, \exp \left(-  S_{\mbox{eff}}(\beta^*,\beta) \right)
\ee
where $ S_{\mbox{eff}}$ is given in Eq.~(\ref{bosaction}).

%\clearpage 


\begin{thebibliography}{11}

\bibitem{Bohr}
A.Bohr,B.R. Mottelson and D.Pines, Phys.Rev.110 (1958) 936

\bibitem{Arim}
F.Iachello and A.Arima, The Interacting Boson Model, Cambridge University Press, Cambridge, 1987

\bibitem{Beli}
S.T.Beliaev and V.G.Zelevinsky, Nucl.Phys.39 (1962) 582

\bibitem{Maru}
T.Marumori, M.Yamamura and A.Tokunaga, Progr.Theor.Phys. 31 ( 1964) 1009; 32 (1964) 726

\bibitem{Otsu}
T.Otsuka,A.Arima and F.Iachello, Nucl.Phys. A309 (1978)1

\bibitem{Otsu1}
T.Otsuka and A.Arima, Phys. Lett.77B (1978)1

\bibitem{Klei}
A.Klein and E.R.Marshalek, Rev. Mod. Phys. 63:375 (1991)

\bibitem{Diep}
A.E.L. Dieperink and G.Wenes, Ann. Rev. Nucl. Part. Sci. 35 (1985) 77

\bibitem{Namb}
Y. Nambu and M. Mukherjee, Phys. Lett. B209 ( 1988) 1; M. Mukherjee and Y. Nambu, Ann. Phys. 191 ( 1991) 143 

\bibitem{Barb}
M.B.Barbaro, A.Molinari, F.Palumbo and M.R.Quaglia, Phys.Rev.C70:034309,2004, nucl-th/0304028

\bibitem{Mira}
V.A.Miransky, Dynamical symmetry breaking in quantum field theories, World Scientific, 1993

\bibitem{Kalo}
M.H.Kalos and P.Whitlock, Monte Carlo Methods, Wiley \& Sons, 1986; A comprehensive account of existing techniques
can be found in: Quantum Monte Carlo Methods in Physics and Chemistry, Edited by M.P. Nightingale and C.J. Umrigar
(Kluwer Academic Publishers, the Netherlands 1999); see also J.A.White, S.E.Koonin and D.J. Dean, Phys. Rev.C 61
(2000) 034303

\bibitem{Cini}
M.Cini and G.Stefanucci, cond-mat/0204311 v1

\bibitem{Palu}
F.Palumbo, Bosonization and Interacting Boson Model,in Proceedings of the 8-th International Spring Seminar on
Nuclear Physics, Paestum 2004, Edited by A.Covello (World Scientific, Singapore, 2005) p.361

\bibitem{Nege}
J. W. Negele and H. Orland, Quantum Many-Particle Systems, Addison-Wesley Publishing Company, 1988

\bibitem{Ripk}
J-P Blaizot and G. Ripka, Quantum Theory of Finite Sysytems, The MIT Press, Cambridge, Massachusetts (1986)

\bibitem{Loiu}
N.Loiudice and F.Palumbo, to be published

\bibitem{Vars}
D.A. Varshalovich, A.N. Moskalev and V.K. Khersonskii, Quantum theory of angular momentum,
World Scientific Publishing Co. Pte.Ltd, 1988


 





\end{thebibliography}
\end{document}